\documentclass[aps,prl,floatfix,twocolumn]{revtex4}
\usepackage{epsfig}
\usepackage{amsmath}
\usepackage{amsfonts}
\usepackage{amssymb}
\usepackage{mathptmx}
\usepackage{eucal}

\DeclareMathOperator{\re}{\mathop{\mathrm{Re}}}
\DeclareMathOperator{\im}{\mathop{\mathrm{Im}}}
\newcommand{\la}{ \left\langle }
\newcommand{\ra}{ \right\rangle }

\begin{document}

\title{Electron sound in metals}

\author{Yu.~A.~Avramenko}
\author{E.~V.~Bezuglyi}
\author{N.~G.~Burma}
\author{V.~D.~Fil'}
\email{fil@ilt.kharkov.ua}
\affiliation{B.Verkin Institute for Low
Temperature Physics and Engineering, 61103 Kharkov, Ukraine}

\begin{abstract}
This paper is devoted to the investigation of electron sound -- oscillations
of the electron distribution function coupled with elastic deformation and
propagating with the Fermi velocity. The amplitude-phase relations
characterizing the behavior of the electron sound in Ga single crystals are
determined experimentally. A model problem of excitation of electron sound in
a compensated metal with equivalent bands is solved for a finite sample with
diffusive scattering of electrons at the interfaces. It was found that the
displacement amplitude of the receiving interface is two orders of magnitude
larger than the elastic amplitude of the wave due to electron pressure. It
was established that the changes occurring in the amplitude and phase of the
electron sound waves at a superconducting transition do not depend on the
path traversed by the wave, i.e. they refer only to the behavior of the
transformation coefficient.
\end{abstract}

\maketitle

\section{I. Introduction}

The existence of different types of waves is characteristic for metals. These
are, first and foremost, oscillations of the electron distribution function
having velocities close to the Fermi velocity. Initially, it was believed
that a necessary condition for their existence is the presence of a magnetic
field \cite{1}, but it subsequently became clear that some types of waves can
also be observed in the absence of a magnetic field. Specifically, the
propagation of zero-sound waves is possible for a certain symmetry of the
Fermi surface (FS) \cite{2}. If regions between which collisional exchange is
impeded for any reason are present on the Fermi surface, then this creates
conditions for the existence of a so-called concentration mode \cite{3,4},
which consists of a periodic redistribution of electrons between these
regions. Actually, the concentration mode is an electron analog of a
first-sound wave. The universal transport mechanism for excitations in a
metal with velocities of the order of the Fermi velocity is a quasiwave
\cite{5}. A quasiwave process is considered to be a purely ballistic effect,
since theoretically it exists in the approximation of noninteracting
electrons. The propagation of bunched waves is possible in metals whose Fermi
surface contains flat sections \cite{6,7}. Since the velocity of all these
waves is close to the maximum Fermi velocity, Landau damping for them, as a
rule, is small and the damping length is of the same order of magnitude as
the electron mean free path. Because of the electron and elastic subsystems
are quite closely coupled, perturbations of the distribution function are
ordinarily accompanied by lattice deformation, i.e. these waves also transfer
elastic deformations with the Fermi velocity. This circumstance is
essentially the basis for the title of the present article, and it provides
the possibility of exciting and detecting them experimentally. These
questions have been discussed in detail theoretically and, for the most part,
confirmed experimentally.

Our motivation for addressing this well-studied problem is as follows. All
theoretical estimates made to date give for the modulus of the coupling
coefficient (i.e. the ratio of the amplitude of the elastic displacement in a
wave to the amplitude of the exciting signal) $K \sim (s/v_F)^2$ ($s$ is the
sound velocity and $v_F$ is the Fermi velocity). However, measurements of the
conversion efficiency in different experiments systematically give values of
$K$ which are several tens of times greater than this estimate. This
discrepancy has already been noted in a study of fast magnetoplasma waves
\cite{8}. A similar picture is also observed for the excitation of zero-sound
waves, concentration modes, and quasiwaves. The present work is devoted to an
analysis of the reasons for such discrepancies for the example of the
excitation of different kinds of waves in ultrapure gallium samples. In the
experimental part a careful determination of the conversion efficiency is
made, the temperature dependencies of the amplitude-phase characteristics of
the elastic disturbances propagating with Fermi velocity are studied, and the
effect of a superconducting transition on them is analyzed.

\begin{figure}[th]
\centerline{\epsfxsize=7.5cm\epsffile{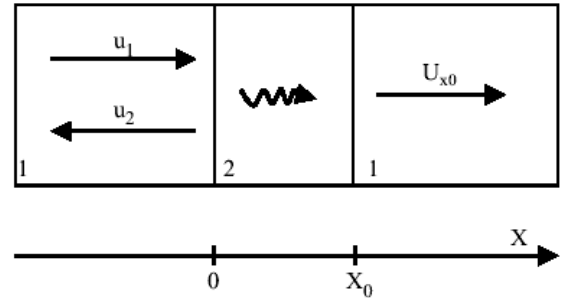}}
\caption{Simplified scheme of the experiment.}
\end{figure}

The theoretical part of this work is computational, i.e. it does not contain
an analysis of different limiting cases, often difficult to implement in
practice, and reduces mainly to a numerical solution of the problem with an
experimentally realizable parameter set. It is devoted to two aspects. As far
as we know, all theoretical estimates of $K$ have been made for a half-space
with a specular boundary. It is naively assumed that the computed amplitude
of the elastic field of the electron sound wave is directly detected by the
receiving piezoelectric transducer. In the present work the problem is solved
for a finite sample with diffusely scattering boundaries. There was an
unqualified expectation that taking account of diffuseness will eliminate the
discrepancies noted above. In any case the calculations of the efficiency of
electromagnetic generation of sound at a diffuse boundary \cite{9} showed
that such a possibility does exist. However, it was found that the
diffuseness of scattering at the boundary increases $K$ by only several
percents. At the same time an analysis of the events on the receiving side
revealed that as a result of the considerable difference between $s$ and
$v_F$ the receiving interface is an effective concentrator of elastic
deformations, increasing by a factor $v_F /s$ compared to the amplitude
computed for the interior volume. In other words, the detected elastic
displacement is determined not by the magnitude of the elastic deformation in
the wave but rather by the much stronger electronic pressure on the
interface.

\section{II. SEVERAL EXPERIMENTAL RESULTS}

A simplified scheme of the experiment performed in the pulsed regime is
presented in Fig.~1. A longitudinal elastic wave with given amplitude $U_1$
is incident on the delay line-sample exciting interface ($x=0$) at $t=0$.
Generally, a reflected wave $U_2$ also exists. The total deformation $U_0=U_1
+U_2$ acting on the interface excites in the sample a fast electron sound
wave and an ordinary sound wave slightly renormalized by the electron-phonon
interaction. The arriving electron sound wave engenders a deformation
$U_{x_0}$ on the receiving interface ($x=x_0$) at times $x_0 /v_F < t < x_0
/s$. The delay lines actually used in the experiment to separate fast signals
from the probing signal do not have ideal acoustic matching with the sample.
However, in the calculations performed in the next section the density of the
delay lines and the sample were assumed to be the same and equal to $\rho$
and the sound velocity in them was also assumed to be the same and equal to
$s$, the sound velocity in the sample, in order to avoid analysis of
reflections, which in the present case are negligible. In this work the
harmonic signals are represented in the form
\begin{equation}
U(t,x) = UK \exp(-\Gamma x) \exp(i\omega t + i\phi),
\end{equation}
where $K$ is a complex transformation coefficient, $\Gamma$ is the damping,
and $\phi$ is the phase of the signal. In a propagating wave $\phi= -k'x$
($k'$ is the real component of the wave number), so that this component of
the phase is always negative. The duration of the leading edge of the
exciting signal ($\sim 0.2$ $\mu$s) is long compared with the period; this
makes it possible to neglect in the theoretical analysis the difference of
the time derivative from $i\omega$. Figure 2a shows the dependence of the
measured amplitude of the fast signals with respect to the excitation
amplitude on the thickness of the sample for different frequencies in the
most interesting case of waves propagation along the [010] axis. The
measurements were performed at temperature $1.7$ K, and the impurity
relaxation time $\tau \sim 10^{-8}$ s. It is evident that the points fall
quite well on a straight line whose slope is frequency-dependent. The
extrapolated straight lines converge at $x_0=0$ practically to a point, which
determines the modulus of the transformation coefficient. Such extrapolation
is valid, generally speaking, only for a single excited signal. In a
two-signal variant this is possible only if the velocity and damping of the
constituent components are close. The frequency dependence of the slopes of
the extrapolated straight lines is likewise close to linear (inset in
Fig.~2a), which is, most likely, an indication that Landau damping is weak.
The intersection of this line with the ordinate determines the
frequency-independent relaxation damping, agreeing well with the known
relaxation time with $v_F \approx 7\times 10^7$ cm/s. The contribution of the
relaxation exponent is presented in Fig.~2a (dotted line).

\begin{figure}[tb]
\centerline{\epsfxsize=8.5cm\epsffile{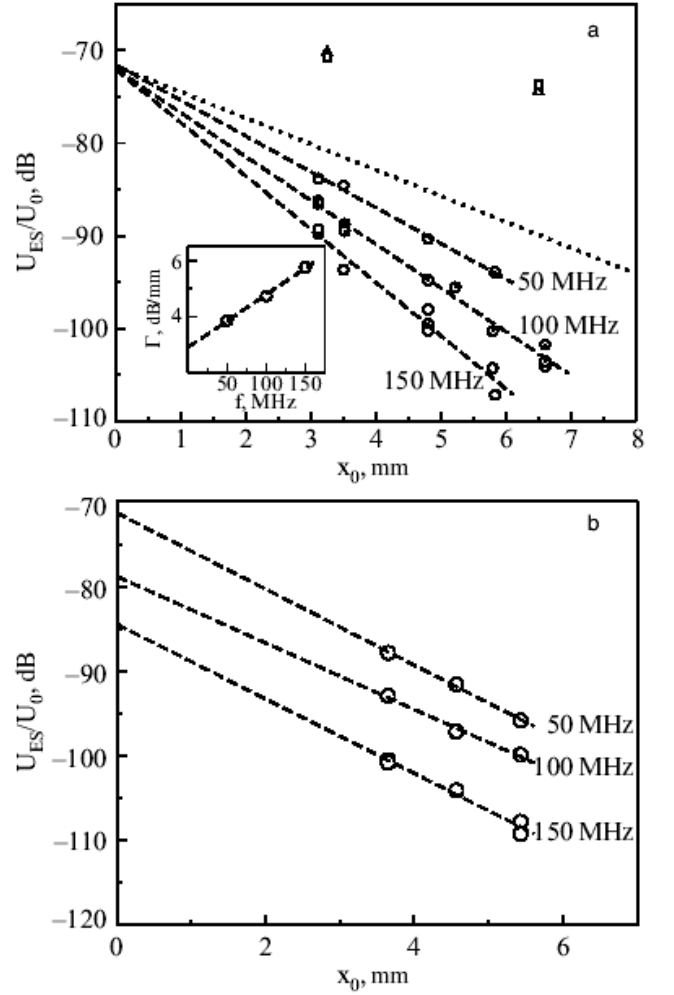}}
\caption{Measured dependencies of the amplitude of electron sound on the
thickness of the sample: $T=1.7$ K, {\bf q}$\|$[010], relaxation damping
(points), calculation for $\nu_1 /\nu_0=0.03$, $F=1$ ($\square$), $\nu_1
/\nu_0=0.03$, $F=0.01$ ($\triangle$) (a). Inset: dependence of the damping on
the frequency. Measured dependencies of the amplitude of electron sound on
the thickness of the sample: $T=1.7$ K, {\bf q}$\|$[100] (b).}
\end{figure}

Similar results for the [100] axis are presented in Fig.~2b. Here, once
again, the thickness dependences can be represented by straight lines.
However, in contrast to the [010] axis the slopes of these straight lines are
practically frequency-independent and are close in magnitude to the slope
determined by the relaxation damping. At the same time the conversion
coefficient determined by the coordinates of the intersection of the straight
lines with the ordinate is frequency-dependent, though the value is close to
that measured on the [010] axis. Apparently, here the situation is close to
the two-signal case with a frequency-dependent phase difference between the
constituent components.

\begin{figure}[tb]
\centerline{\epsfxsize=8.5cm\epsffile{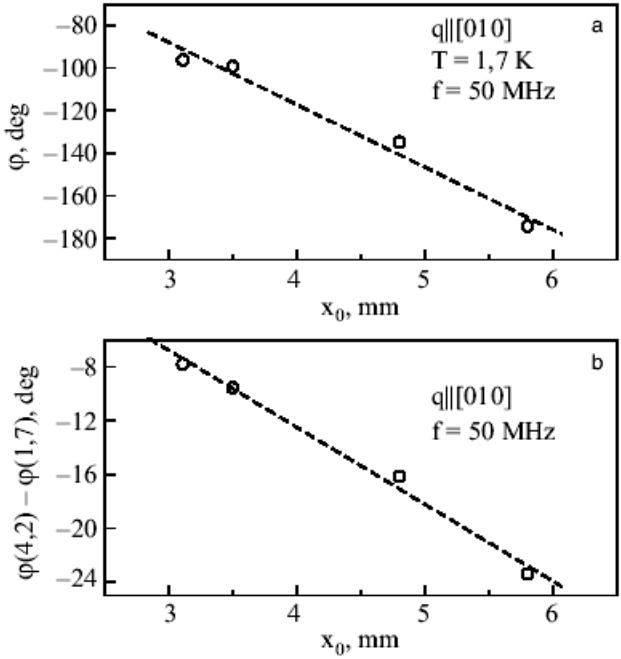}}
\caption{Measured sample thickness dependences of the phase: total change of
the phase (a); the phase increment accompanying a temperature change from 1.7
to 4.2 K (b).}
\end{figure}

We shall now discuss the phase characteristics of the signals. It is hardly
possible to determine the reference point experimentally with adequate
accuracy. For this reason only relative measurements are meaningful. Figure
3a shows the thickness dependence of the phase which the electron sound wave
acquires as it propagates along the [010] axis. A straight line also fits it
well, its slope giving a wave velocity close to $7\times 10^7$ cm/s at
$T=1.7$ K. The phase of the signals decreases with increasing temperature;
this indicates either a change of the velocity of the electron sound wave or
a strong dependence of the phase of the transformation coefficient on the
relaxation time. It has been shown in a previous work \cite{4} that a
substantial decrease of the velocity is possible only in a two-band model of
the spectrum with strongly impeded interband scattering (to a level of
several percent of the intraband scattering), which, at first glance,
appeared to be quite unusual. One of our tasks in the present work was to
separate two possible contributions to the change of the total phase of the
electron sound wave. Figure 3b shows the thickness dependence of the
temperature increments to the phase. Evidently, the straight line fit has a
nonzero slope, indicating that the wave velocity decreases by 20\%. At the
same time, the nonzero coordinate of intersection of the straight line and
the ordinate likewise attests to a simultaneous small increment to the phase
of the transformation coefficient.

More detailed information can be obtained from the temperature dependences of
the phase of the signal in samples of different length. Figure 4a shows the
phase changes measured from the level at $T=1.7$ K ([010] axis). The change
of the electron sound velocity can be found by representing the signals in
the form (1). A decrease of $\tau$ to $10^{-9}$ s ($T \sim 5$ K) results in a
gigantic (40\%) decrease in the velocity (Fig.~5a). The phase of the
transformation coefficient increases somewhat as $\tau$ decreases (Fig.~5b).
The amplitude dependences are presented in Fig.~4b. Analysis shows that the
change in modulus of the transformation coefficient in the limit of the
measurement error is negligible.

\begin{figure}[tb]
\centerline{\epsfxsize=8.5cm\epsffile{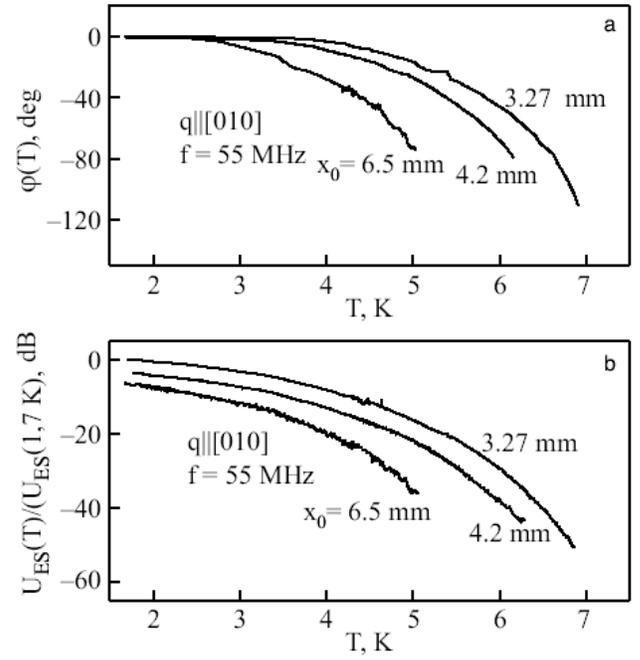}}
\caption{Temperature dependences of the amplitude-phase characteristics of
electron sound: phase change measured from 1.7 K (a); amplitude change (the
curves are shifted by 4 dB) (b).}
\end{figure}
\begin{figure}[tb]
\centerline{\epsfxsize=8.5cm\epsffile{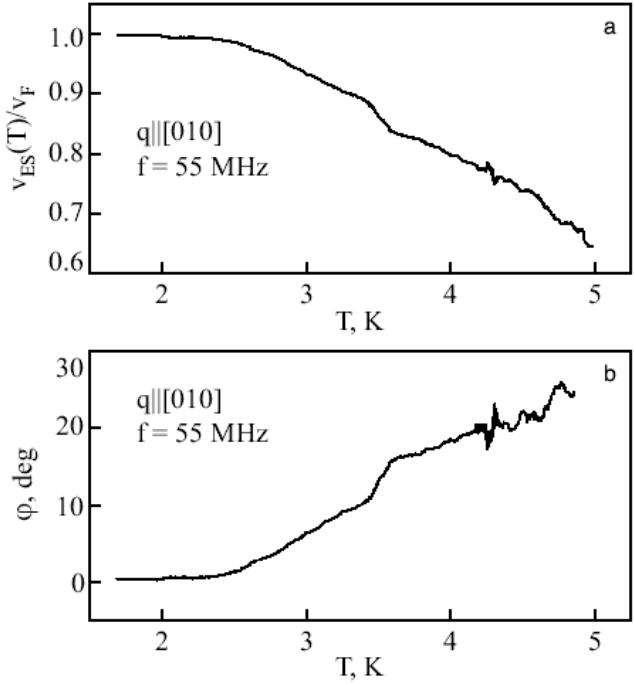}}
\caption{Temperature dependence of the electron sound velocity (a) and of the
phase of the transformation coefficient (b).}
\end{figure}

The phenomena described above are absent for waves propagating in the [100]
direction -- the phases remain practically unchanged right up to vanishing of
the signals.

In concluding this section we wish to emphasize the essential experimental
facts: 1. the modulus of the transformation coefficient at T=1.7 K ($\tau
\sim 10^{-8}$ s) lies in the range $-70 \ldots -80$ dB; 2. the velocity of
electron sound waves propagating in the [010] direction depends strongly on
the relaxation rate, dropping by approximately 40\% at $\tau \sim 10^{-9}$ s;
and 3. the phase of the transformation coefficient increases somewhat ($\sim
20$ deg) with increasing scattering.

\section{III. THEORETICAL ANALYSIS}

The one-dimensional kinetic equation, describing the process, for the
nonequilibrium correction $\psi(x)$ to the distribution function
$n=n_F(E)+(\partial n_F /\partial E)\psi$ ($n_F(E)$ is the Fermi function of
the total quasiparticle energy $E$) has the form
\begin{equation}
i\omega(1+\hat{f})^{-1}\psi + v\frac{\partial}{\partial x} (\psi -e\phi) +
i\omega\Lambda \frac{\partial U}{\partial x} = I(\psi).
\end{equation}
Here $v =v_{Fx}$ and $\Lambda =\Lambda_{xx}$ are the longitudinal components
of the electron velocity and the deformation potential, $\phi$ is the
electric potential, $\hat{f}$ is the Fermi-liquid interaction (FLI) operator,
$U$ is the elastic deformation, and $I(\psi)$ is the collision integral which
takes account of the law of conservation of the number of particles during
scattering \cite{10}
\begin{equation}
I_\alpha(\psi) = \sum\nolimits_\beta w_{\alpha\beta} \left( \la \psi
\ra_\beta - N_{F\beta} \psi_\alpha \right).
\end{equation}
Here and below Greek indices enumerate the electronic groups,
$w_{\alpha\beta} =w_{\beta\alpha}$ is the scattering (assumed to be isotropic
for simplicity) probability matrix, the brackets signify integration over the
FS in the $\alpha$th zone
\begin{equation} \nonumber
\la\mu \ra_\alpha \equiv \frac{2}{(2\pi\hbar)^3} \int_{{\bf{p}}\in \alpha}
\frac{dS_{\bf p}}{v_F} \mu({\bf p}),
\end{equation}
and $N_{F\alpha} =\la 1\ra$ is the density of states in the $\alpha$th zone.
In the model examined below, which takes account of only the isotropic part
of Landau correlation function, the action of the operator $\hat{f}$ on an
arbitrary function $\mu({\bf p})$ of the quasimomentum {\bf p} in the
$\alpha$th zone is determined by the relation
\begin{equation} \nonumber
(\hat{f} \mu)_\alpha = (\hat{f}\la \mu\ra)_\alpha \equiv \sum\nolimits_\beta
f_{\alpha\beta} \la \mu\ra_\beta,
\end{equation}
where $f_{\alpha\beta} = f_{\beta\alpha}$. is the matrix of the coefficients
of the FLI.

To find a relation between the electric and elastic fields the kinetic
equation (2) must be supplemented by the electroneutrality condition
\begin{equation} \nonumber
\sum\nolimits_\alpha \la(1+\hat{f})^{-1} \psi\ra_\alpha =0,
\end{equation}
which in the geometry under consideration reduces to the condition for the
absence of a longitudinal current
\begin{equation}
\sum\nolimits_\alpha \la v\psi\ra_\alpha =0,
\end{equation}
as well as an equation from the theory of elasticity that takes account of
the contribution of the force $f_e$ exerted by the electrons on the vibrating
lattice \cite{11,12}:
\begin{align} \nonumber
&\omega^2 U = - s^2\frac{\partial^2 U}{\partial x^2} + f_e,
\\
&f_e = \rho^{-1} \frac{\partial}{\partial x} \sum\nolimits_\alpha \la \Lambda
(1+\hat{f})^{-1} \psi \ra_\alpha \equiv \rho^{-1} \frac{\partial W}{\partial
x}.
\end{align}

In what follows, as in Ref.~\onlinecite{3} and \onlinecite{4}, we shall
examine a simplest model of a compensated metal with equivalent $e$ and $h$
spherical Fermi surfaces ($v_{Fe} = v_{Fh}$, $N_e = N_h$, and $\Lambda_e =
-\Lambda_h$) with the difference between the intra- and interband FLI
coefficients and the corresponding scattering rates being preserved. It is
obvious that for such a symmetry of the problem $\psi_e = -\psi_h$ and
$\phi=0$. As a result, it is sufficient to study the solution of the kinetic
equation for a single sheet of the FS, doubling $f_e$ in the elasticity
equation (5). In addition, since the value of the nonrenormalized deformation
potential averaged over both bands is zero, we have $\Lambda_{xx} = mv^2_x$.
We introduce the following notations:
\begin{align} \nonumber
&F_{ee}=F_{hh}=F_0, \quad F_{eh} = F_1, \quad F = F_0-F_1, \quad
F_{\alpha\beta} = N_F f_{\alpha\beta},
\\
&\nu_{ee} = \nu_{hh} = \nu_0, \quad \nu_{eh} = \nu_1, \quad \nu_{\alpha\beta}
= N_F w_{\alpha\beta}.
\end{align}
Taking account of the simplifications following from the model the kinetic
equation becomes
\begin{subequations}
\begin{align} \nonumber
&i\omega (1+\hat{f})^{-1} \psi + v\frac{d\psi}{dx} = -i\omega\Lambda
\frac{dU}{dx} -\nu^+ \psi +\nu^- \frac{\la \psi \ra}{N_F}
\\
&\nu^\pm = \nu_0 \pm \nu_1.
\end{align}
\end{subequations}
Applying the operator $(1+\hat{f})$ to the left hand side, we arrive at the
equation
\begin{align} \nonumber
&i\tilde{\omega} \psi + v\frac{d\psi}{dx} = -i\omega\Lambda \frac{dU}{dx}
-\nu^+ \psi +\nu^- \la \psi\ra\frac{\omega^-}{N_F} \equiv A(x),
\\\nonumber
&i\tilde{\omega} = i\omega + \nu^+, \quad \omega^- = \nu^- + \frac{i\omega
F}{1+F} \equiv \nu^- +i\omega F_2. \qquad\quad \text{(7b)}
\end{align}
The solution of Eq.~(7b) in the coordinates of Fig.~1 has the form
\begin{subequations}
\begin{align}
&\psi_{v>0} = Ce^{-Lx} + e^{-Lx}\int_0^x \frac{A(x')}{v}e^{Lx'}\,dx', \quad
L= \frac{i\tilde{\omega}}{v},
\\
&\psi_{v<0} = C_1 e^{-L(x-x_0)} + e^{-Lx}\int_{x_0}^x
\frac{A(x')}{v}e^{Lx'}\,dx'.
\end{align}
\end{subequations}

The solution (8a) and (8b) is suitable for any kind of the scattering
boundary. For diffusive scattering, $C_1$ and $C$ are ``true'' constants,
while for specular scattering they are functions of $v$. We shall assume that
the exciting boundary ($x =0$) is diffusive (i.e. $C$ is a constant), and for
now we shall not specify the type of the reflection at the receiving
interface.

The average value $\la \psi \ra$ can be represented in the form \cite{13}
\begin{align}\nonumber
\la\psi\ra &= C\la e^{-Lx}\ra_> -\la C_1 e^{-L(x-x_0)} \ra_<
\\
&+ \la \int_0^{x_0} \frac{A(x')}{v} e^{-L|x-x'|} dx' \ra.
\end{align}
The subscripts indicate the part of the Fermi sphere (the sign of the
longitudinal component of the Fermi velocity) over which the averaging is
performed; the absence of a subscript means that the averaging is performed
over the entire zone. We shall find $W$ by averaging the expression (7a) with
the weight $\Lambda$ and using (7b) to determine $\la \Lambda v d\psi/dx\ra$:
\begin{align}\nonumber
\frac{W}{2} &= \la \Lambda\psi \ra -F_2 \la \Lambda\ra \frac{\la\psi\ra}{N_F}
= C\la \Lambda e^{-Lx}\ra_> +\la C_1 \Lambda e^{-L(x-x_0)}\ra_<
\\
&+\la \Lambda\int_0^{x_0}\frac{A(x')}{v} e^{-L|x-x'|}dx'\ra_> -F_2
\la\Lambda\ra \frac{\la\psi\ra}{N_F}.
\end{align}

We shall solve the problem in the Fourier representation by the Wiener-Hopf
method \cite{14}, assuming all physical fields outside the sample to be zero.
To use the properties of a convolution, we extend the definition the
integrals on the right-hand sides of Eqs.~(9) and (10) over the entire $x$
axis, adding and subtracting the corresponding functions
\begin{align}\nonumber
\la\psi\ra_k &\equiv \int_0^{x_0}\la\psi\ra e^{-ikx}dx =
C\la\frac{1-e^{-(L-ik)x_0}}{L+ik}\ra_>
\\
&- \la \frac{C_1(e^{-ikx_0} - e{-Lx_0})}{L+ik}\ra_< +\la\frac{A_k}{v}
\frac{2L}{L^2+k^2}\ra_> \nonumber
\\
&-\la\frac{\alpha_1}{v(L-ik)}\ra_> - \la\frac{\alpha_2
e^{-(L+ik)x_0}}{v(L+ik)}\ra_>, \nonumber
\\
\alpha_1 &= \int_0^{x_0} A(x')e^{-Lx'} dx', \quad \alpha_2 =\int_0^{x_0}
A(x')e^{Lx'} dx',
\end{align}
where $A_k$ is the Fourier component of $A(x)$.

The last two terms in Eq.~(11) are analytic for $k'' = \im k > -\nu^+/v$ and
$k'' < \nu^+/v$, respectively:
\begin{align}\nonumber
\frac{W_k}{2}&= C\la\frac{\Lambda(1-e^{-(L+ik)x_0})}{L+ik}\ra_> -
\la\frac{C_1 \Lambda(e^{-ikx_0}-e^{-Lx_0})}{L+ik}\ra_<
\\ \nonumber
&+\la \frac{A_k}{v}\frac{2\Lambda L}{L^2+k^2}\ra_>
-\la\frac{\Lambda\alpha_1}{v(L-ik)}\ra_>
\\
&-\la\frac{\Lambda \alpha_2 e^{-(L+ik)x_0}}{v(L+ik)}\ra_> -F_2 \la\Lambda\ra
\frac{\la\psi\ra_k}{N_F}.
\end{align}
The Fourier transform of the elasticity equation has the form
\begin{align}
&i(q^2-k^2)U_k=(k-q)U_{x_0}e^{-ikx_0}-(k+q)U_0+2qU_1-\frac{kW_k}{\rho s^2},
\end{align}
where $q=\omega/s$ is the wavenumber of the sound. The relation (13) was
obtained using the boundary conditions for the displacement $U_0=U_1+U_2$ and
for the elastic stress
\begin{subequations}
\begin{align}
-iqU_1+iqU_2 = \frac{dU_0}{dx} -\frac{W(0)}{\rho s^2} \quad (x=0),
\\
-iqU_{x_0} = \frac{dU_{x_0}}{dx} - \frac{W(x_0)}{\rho s^2} \quad (x=x_0).
\end{align}
\end{subequations}

We call attention to the fact that the derivatives on the left and right hand
sides in the boundary conditions are calculated on different sides of the
interface, and they are by no means equal. Using the identities
\begin{align} \nonumber
&\la\frac{\Lambda}{L+ik}\ra = \epsilon\la \frac{L-ik}{L^2} -
\frac{1}{L+ik}\ra,\;\; \epsilon =\frac{E_F k_0^2}{k^2},
\\ \nonumber
&\la\frac{\Lambda L}{L^2+k^2}\ra = \epsilon\la \frac{1}{L} -
\frac{L}{L^2+k^2}\ra, \;\;  k_0 = \frac{i\widetilde{\omega}}{v_F},\;\;
E_F=mv_F^2,
\end{align}
we obtain
\begin{align} \nonumber
\frac{W_k}{2} &= -\epsilon\Bigl[\la\psi\ra_k T(k) +I_0(k) +I_1(k)e^{-ikx_0}
+2\la\Lambda\ra \frac{\omega}{\widetilde{\omega}}
\\ \nonumber
\times \biggl(&U_{x_0} e^{-ikx_0} -U_0+ikU_k\biggr)\Bigr], \quad T(k) =
1-\frac{\omega^-}{\widetilde{\omega}} +\frac{F_2}{3}\frac{k^2}{k_0^2},
\\ \nonumber
I_0(k) &= C\la \frac{L-ik}{L^2}\ra_> +\la\frac{C_1(L-ik)}{L^2}\ra_<
-\la\frac{\alpha_1(L+ik)}{vL^2}\ra_>,
\\ \nonumber
I_1(k) &= -C\la \frac{e^{-Lx_0}(L-ik)}{L^2}\ra_>
+\la\frac{C_1(L-ik)}{L^2}\ra_<
\\
&-\la\frac{\alpha_2 e^{-Lx_0}(L-ik)}{vL^2}\ra_>,
\end{align}

Substituting the expressions (15) into the elasticity equation and
transferring to the left-hand side the terms with $U_k$ and $\la\psi\ra_k$
and all terms containing $e^{ikx_0}$ we obtain
\begin{align} \nonumber
&k\Delta(k)U_k -\frac{i\sigma k_0^2}{n} T(k)\la\psi\ra_k
+\Bigl[-ik(k-q)U_{x_0} +\frac{ik_0^2}{n}I_1(k)
\\ \nonumber
&+\sigma k_F k_0U_{x_0}\Bigr]e^{-ikx_0} = -ik(k+q)U_0 +2ikqU_1
\\
&-\frac{i\sigma}{n} k_)^2I(k) +\sigma k_F k_0U_0, \quad \Delta(k) = k^2-q^2
+i\sigma k_F k_0
\end{align}
where $k_F = \omega/v_F$ should not be confused with the Fermi momentum,
$\sigma = E_F/Ms^2$, $M$ is the ion mass, and $n$ is the carrier density in
the zone. The parameter $\sigma$ in the free-electron model is close to 1 but
in a real case, because of the electron-phonon renormalization of the Fermi
surface, as a rule, it is less than 1. In what follows $\sigma=0.25$ is used
in numerical estimates.

The function on the right-hand side of Eq.~(16) is analytic on the entire
complex plane and by Liouville's theorem can be represented as a quadratic
polynomial:
\begin{align} \nonumber
P_2(k) &= \alpha k^2 + \beta k +\gamma,
\\ \nonumber
\alpha &= -iU_0, \quad \gamma = \sigma k_F k_0 U_0 -\frac{i\sigma k_0^2}{n}
I_0(0).
\end{align}

The left-hand side of Eq.~(16) contains terms which increase exponentially as
$k'' \to -\infty$. Evidently, in order for the relations (16) to hold $U_k$
and $\la\psi\ra_k$ finite, they must contain components that exactly
compensate this growth. Then the relation (16) can be split into two
independent equations and each equation can be solved separately. Such
splitting makes it possible to circumvent the inconveniences due to the
absence of singular points in the Fourier transforms taken on a finite
interval and to make use of contour integration to obtain the inverse
transform. Likewise, the physical meaning of the terms containing the factor
$\exp(-ikx_0)$ is obvious: when the inverse transformation is computed the
contour integrals with the factor $e^{-ik(x_0-x)}$ ($x<x_0$) must be closed
in the bottom half-plane in $k$ space; this corresponds to waves propagating
in the negative $x$ direction.

Eliminating $\la\psi\ra_k$ for forward waves from Eqs.~(11) and (16) we
obtain
\begin{align} \nonumber
&Z(k)[kB_z(k)U_k -V_0(k)] = 3\sigma k_0^2 T(k) d\Bigl[ k_Fk_0 q^2 U_0
\\ \nonumber
&+k\Delta(k) C +ikk_Fk_0\beta +\frac{k_Fk_0^3}{n}I_0(0)\Bigr] +\frac{i\sigma
k_0^2}{n} T(k)B_z(k) J_0(k),
\\ \nonumber
&V_0(k) = P_2(k)\frac{\omega^-}{i\widetilde{\omega}} k^2 -3\sigma k_Fk_0^3
T(k)U_0 -3\sigma kk_0^2 T(k) \frac{C}{E_F},
\\ \nonumber
&B_z(k) =\frac{\omega^-}{i\widetilde{\omega}} k^2\Delta(k)-3i\sigma k_Fk_0^3
T(k),
\\ \nonumber
&J_0(k)= C\la\frac{1}{L-ik}\ra_> + \la\frac{C_1 e^{-Lx_0}}{L+ik}\ra_< - \la
\frac{\alpha L}{v(L-ik)}\ra_>,
\\
&Z(k) = d\Delta(k)+B_z(k)\la\frac{1}{L^2+k^2}\ra_>, \quad d =
1-\frac{\omega^-}{i\widetilde{\omega}}.
\end{align}

The characteristic equation $Z(k)=0$ determines the wave\-num\-bers of the
propagating waves. In general, there are two pairs of roots: $r_1 \approx \pm
k_0$ and $r_2 \approx \pm q$ -- sound slightly renormalized by the
electron-phonon interaction. In addition, there are two branch points $k= \pm
ik_0$ with which the quasiwave process is associated.

Let the roots $r_1$ and $r_2$ ($\re r_{1,2} < 0$, $\im r_{1,2}> 0$) lie in
the upper half-plane. According to the Wiener-Hopf method, the function
$Z(k)$ must be factorized, i.e. represented as a product of two functions
which are analytic, respectively, in the upper and lower half-planes and have
a common strip of analyticity. The function $Z(k)$ has no singularities near
the real axis, including $k=0$. As $k \to \infty$, $Z(k)$ behaves as $k^2$.
We introduce the function
\begin{align} \nonumber
\widetilde{Z}(k) = \frac{Z(k)}{(k+r_1)(k+r_2)},
\end{align}
which can be factorized according to a general rule \cite{14}, $Z(k)=
\tau^+(k) \tau^-(k)$, where
\begin{align}
\tau^+(k) =\exp\left(\frac{1}{2\pi i}\int_{-\infty+i\delta}^{+\infty+i\delta}
\frac{\ln \widetilde{Z}(\xi)}{\xi -k} d\xi\right).
\end{align}
Since the poles of $Z(k)$ in the function $\widetilde{Z}(k)$ for $k''<0$ have
already been eliminated, the calculation of the integral (18) by closing the
contour in the lower half-plane reduces to the contribution of only the cut
$C'$, extending from the point $ik_0$ to infinity. To choose the cut $C'$ it
is necessary to make sure that the integrand does no cross the cut itself,
accompanied by a jump of the imaginary part of the logarithm. Specifically,
the commonly used variant where the cut is drawn along a ray emanating from
the coordinate origin and passing through the branch point does not work in
the present case. Any cut made near the imaginary axis will work.

For waves propagating in the forward direction the amplitude behaves as
$e^{-\Gamma x}$ ($\Gamma >0$), so that $U_k$ is analytic for all $k'' <
\Gamma$. Dividing (17) by $(k+r_1)(k+r_2)\tau^+(k)$, we obtain the functional
equation
\begin{widetext}
\begin{equation}
\tau^-(k)[kB_z(k)U_k-V_0(k)] = -\frac{3\sigma k_0^2 T(k)dk_F k_0 q^2 U_0
+k\Delta(k)C + ik_F k_0 k\beta +\frac{k_F
k_0^3}{k}I_0(0)}{(k+r_1)(k+r_2)\tau^+(k)} + \frac{i\sigma}{n} \frac{k_0^2
T(k)B_z(k)J(k)}{(k+r_1)(k+r_2)\tau^+(k)}.
\end{equation}
\end{widetext}

The left-and right-hand sides of Eq.~(19) are analytic, respectively, in the
lower and upper half-planes, they possess a common strip of analyticity near
the real axis, and they can be represented as cubic polynomials
\begin{align} \nonumber
P_3(k) = T(k)(A_1 k +A_0).
\end{align}
Setting $k=0$ and using $I_0(0)=J_0(0)$ we obtain
\begin{align}
A_0 = -\frac{2\sigma k_F k_0^3 d}{r_1 r_2 \tau^+(0)} q^2 U_0.
\end{align}
We find the coefficient $A_1$ as well as the unknown constants $C$, $\beta$,
and $I_0(0)$ by setting $k$ equal to the roots $b_i$ of the quartic
polynomial $B_z(k)$.

We call attention to a paradoxical, at first glance, circumstance. Even
though $x_0$ appears explicitly in $I_0(0)$, this constant does not depend on
$x_0$, just as it does not depend on $C_1$ or on the character of the
reflections at the receiving interface. The situation is that these
quantities also appear in $\alpha_1$, since as one can see from Eq.~(11),
$\alpha_1$ is the $(-iL)$th Fourier component of the perturbing field $A(x)$,
which is determined by the contribution of the forward and backward waves.
The independence of $I_0(0)$ from $x_0$ and $C_1$ shows that the contribution
of the backward wave to $I_0(0)$ is completely annihilated. Of course, this
does not mean that the field of the forward wave is absolutely insensitive to
the distance up to the receiving interface and the character of the
reflections from it. In reality, the amplitudes of the propagating waves are
determined in terms of the field on the emitting boundary $U_0$ and,
naturally, the backward wave contributes to it. But, as will become clear
below, this correction is $\sim (s/v_F)^2$ for a diffuse boundary at $x=x_0$
and, apparently, $\sim s/v_F$ for a specular boundary.

Thus the elastic field of the forward wave at any point of
the sample is practically independent of the thickness of the
latter and can be calculated using the relations obtained for a
half-space. The Fourier transform of the elastic component
of the field of the forward wave has the form
\begin{subequations}
\begin{align}
&\!\!\!\!\!U_{k1} = \frac{T(k)(A_1k+A_0)(k+r_1)(k+r_2)\tau^+(k)}{kB_z(k)Z(k)}
+ \frac{V_0(k)}{kB_z(k)}.
\end{align}
\end{subequations}
In Eq.~(21a) $\tau^-(k)$ is replaced by
\begin{align} \nonumber
\frac{Z(k)}{(k+r_1)(k+r_2) \tau^+(k)}.
\end{align}

It is easy to show that $k=0$ and $k=b_i$ are not poles of $U_{k1}$ and the
physical fields are determined only by the roots and branch point of $Z(k)$.
The amplitude of the sound wave is practically equal to $U_0$, while the
electron sound wave of interest to us is the total contribution of the mode
with $k =r_1$ (if this root exists) and a quasiwave.

The Fourier transform of the elastic component for the half-space with a
specular interface boundary is calculated in Refs.~\onlinecite{3} and
\onlinecite{4}. In our notation it has the form
\begin{align}\nonumber
U_{k1} = -\frac{6\sigma k_F k_0^3 T(k)dq^2 U_0}{kB_z(k) Z(k)}. \qquad
\qquad\qquad \qquad\qquad \text{(21b)}
\end{align}
The terms eliminating the singularities at $k=0$ and $k=b_i$ have been
dropped in Eq.~(21b).

\begin{figure}[tb]
\centerline{\epsfxsize=8.5cm\epsffile{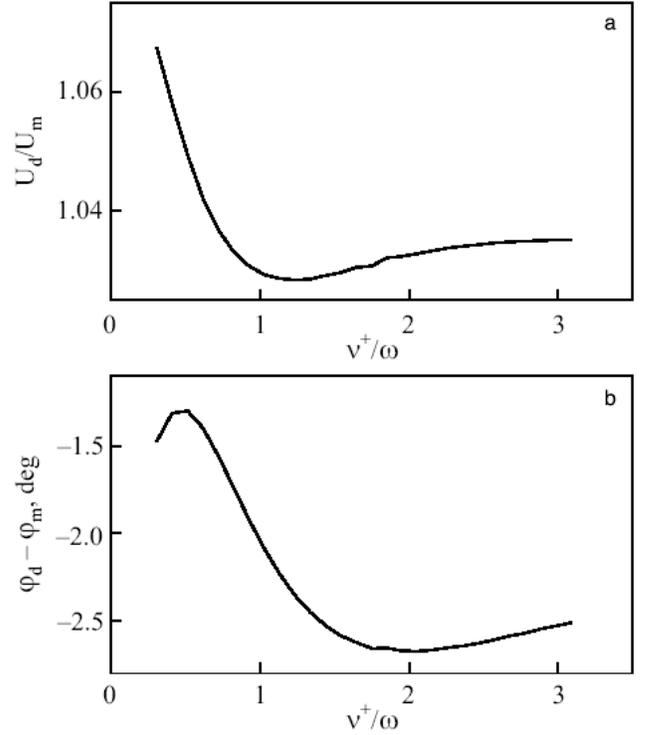}}
\caption{Comparison of the results for diffuse (d) and specular (m)
boundaries: $\nu_1 /\nu_0=0.03$, $F=0.6$; the quantity $k_F x_0=1.2$,
corresponding to $x_0 \sim 0.33$ cm, was used in the calculation; amplitude
ratio (a) and phase difference (b).}
\end{figure}

The term with $A_0$ makes the main contribution to the polynomial $P_3(k)$ at
$k \sim k_0$, so that both solutions (21a) and (21b) are practically
identical even in the literal form. Figures 6 shows for a concrete parameter
set the ratio of the moduli of the signals and their phase difference, which
are calculated using the relations (21a) and (21b). It is evident that even
diffuseness results in a very small increase of the transformation
efficiency. But the magnitude of the total signal itself, calculated, for
example, for $x_0=3$ mm, at $T =1.7$ K and $v_F /s$=200 is $-110$ dB, which
is much less than the experimental values (see Fig.~2).

We shall now discuss the events which occur on the receiving interface. If
the reflections on it are diffuse, the constant $C_1$ is independent of $v$
and can be removed from the averaging operation. The functional equation
determining the Fourier transform of the backward waves is identical to, with
the exception of the common factor $\exp(-ikx_0)$, some relabeling, and
mutual substitutions of the regions of analyticity, the expression (19) with
$-U_{x_0}$ replacing $U_{x_0}$. This means that the as yet unknown elastic
displacement $U_{x_0}$ arising at the interface engenders backward waves with
the same efficiency as the initial perturbations at $x=0$. Specifically, the
main contribution to the waves moving away from the boundary will be acoustic
with amplitude practically the same as $U_{x_0}$. The relation between $U_0$
and $U_{x_0}$ is determined by the condition of continuity of the elastic
stresses on both sides of the interface (14b). The derivative $dU_{x_0} /dx$
can be replaced by $-ir_2 U_{x_0} \approx iqU_{x_0}$, and only the arriving
electron sound wave need be taken into account in the electron-elastic
potential $W(x_0)$. According to Eq.~(13) the Fourier transform $W_k$ has the
form
\begin{align}
-\frac{W_k}{\rho s^2} = \frac{(k^2-q^2)U_{k1}}{ik}.
\end{align}

All terms which are unrelated to the singular points of $Z(k)$ have been
dropped in Eq.~(22). Using the representation of $U_{k1}$ found above, the
inverse Fourier transform makes it possible to find $W(x_0)$. Since the
characteristic $k \approx k0 \ll q$ in the electron sound wave incident on
the boundary, we obtain for the amplitude of the elastic field
\begin{align} \nonumber
U_{x_0} \sim \frac{v_F}{2s} U_f \approx \frac{s}{2v_F} U_0,
\end{align}
where $U_f$ is the displacement amplitude in the incident wave.

We arrive at the following conclusion, which is the essential result of the
present work -- the experimentally recorded displacements engendered by the
electron sound wave are determined by, first and foremost, the electronic
pressure on the boundary and are substantially greater than the elastic
component of the wave itself. Although this result was obtained for a
specific model of a compensated metal with equivalent bands, its validity
evidently is not limited by the properties of this particular model.

In the case of specular reflection, aside from the incident electronic waves,
reflected waves with amplitude comparable to the first waves and also
contributing to $W(x_0)$ also appear. This complicates the solution of the
problem. However, if the structure of the elastic field arising in the metal
is ignored and only the magnitude of the signal on the receiving interface is
considered, then we can use the reciprocity principle \cite{15}. According to
this principle, if the transmitting medium is not gyrotropic and no linear
effects occur in it, then the recorded amplitude is independent of which
piezoelectric transducer is the source of the perturbation and which is the
receiver. Since, as the preceding discussion has made clear, the scale of the
events at a diffuse boundary is independent of the specularity of the
generating interface, exactly the same result must obtain in the opposite
case.

We shall now briefly discuss the qualitative results following from the
calculations performed above and from their comparison with experiment.
Figures 7 and 8 demonstrate the typical computed dependences of the amplitude
and phase of the elastic displacement, recorded on the receiving interface
for the cases of weak and intense interband scattering. The results are
presented in the interval of total relaxation rates which is close to the
experimental interval. It was assumed in the calculations that the ratio
$\nu^- /\nu^+$ is independent of the temperature. The contribution of the
pole (if it exists), the contribution of the branch point, and the total
signal were presented separately. The contribution of the pole is usually
identified with zero sound in the limit of weak scattering and with the
concentration mode in the opposite case \cite{3,4}. In what follows, for
brevity, this contribution will be called the wave component. The
contribution of the branch point will be called the quasiwave part, which
conforms to the accepted terminology \cite{5}. The interband scattering
parameter (Figs.~6 and 7) was chosen, as done previously in
Ref.~\onlinecite{4}, so that it would describe the actually observed decrease
of the electron sound velocity ${\bf q} \| [010]$. The essential conclusions
following from the calculations performed reduce to the following.

\begin{figure*}[bt]
{\epsfxsize=17cm\epsffile{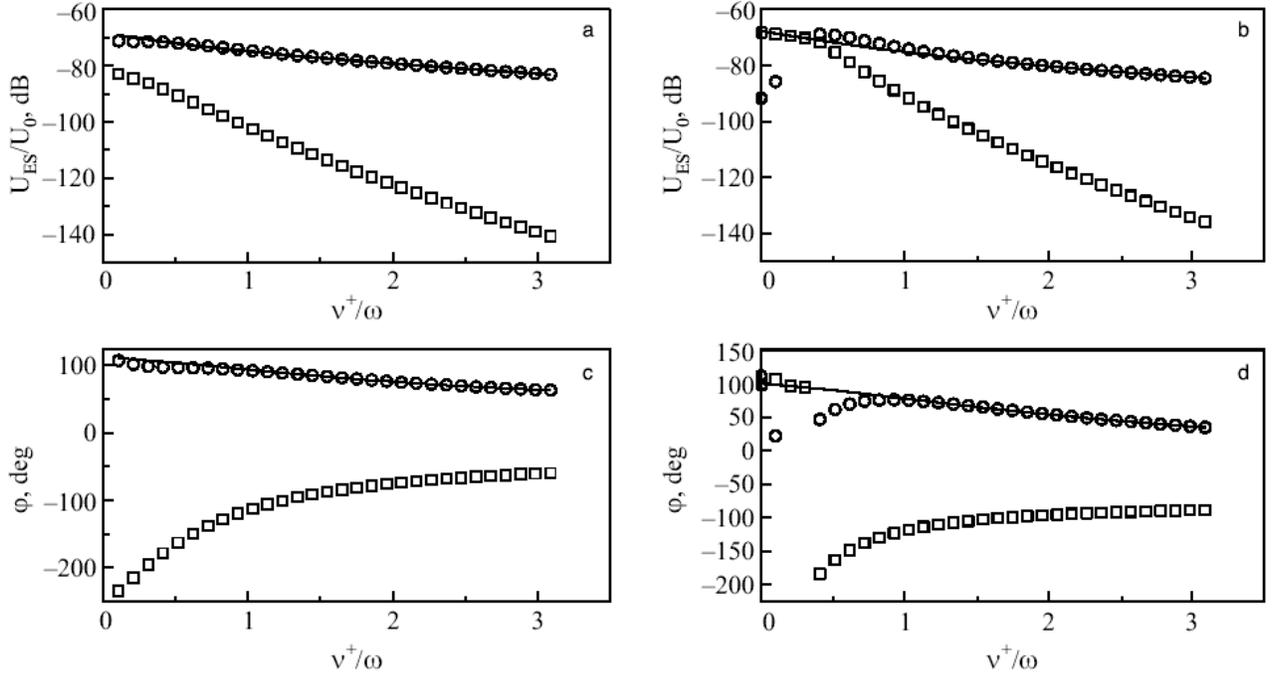}}
\caption{Computed dependences of the amplitude and phase of electron sound
with weak interband exchange $\nu_1 /\nu_0=0.03$ versus the scattering
parameter $k_F x_0=1.2$: wave component ({\Large $\circ$}), quasiwave
component ($\square$), total signal (line); amplitude and phase for $F=1$
(a,c) and $0.3$ (b,d).}
\end{figure*}

1. For a given interband scattering rate the amplitude of the total signal is
practically independent of the FLI parameter (see also Fig.~2). At the same
time a change of $F$ results in a redistribution of the intensities between
the wave component and the quasiwave. For weak interband exchange in the
region of existence of the concentration mode (high temperatures) the latter
always dominates with the exception of $\nu^+$ adjoining the boundary of its
existence. The wave component also predominates for low temperatures and $F
\geqslant 0.6$. The quasiwave predominates for intense interband exchange and
small $F$, but for $F \sim 1$ and small $\nu^+$ the zero-sound solution makes
the main contribution. There exists a theorem asserting that in the
degenerate case of the type studied in the present work and in the absence of
scattering zero sound exists for arbitrarily weak FLI \cite{2} The result
presented in Fig.~7b agrees with this assertion, but the amplitude of the
wave component is small and is completely masked by the quasiwave.

\begin{figure*}[bt]
{\epsfxsize=17cm\epsffile{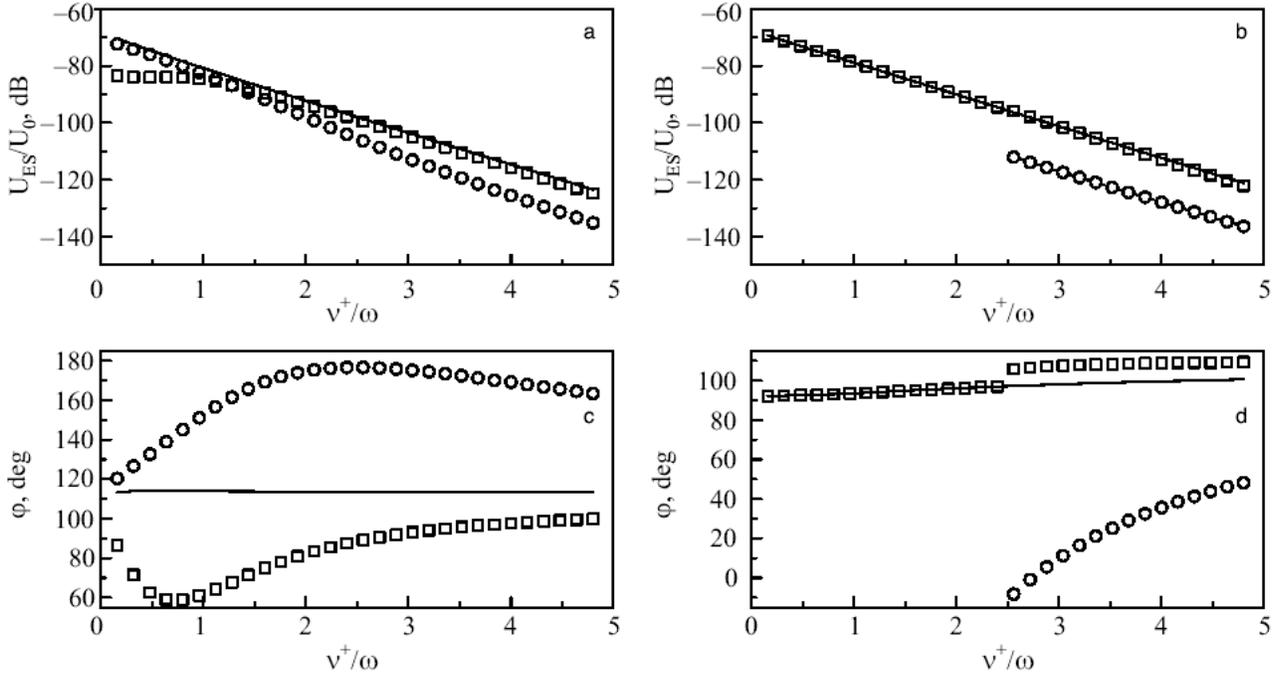}}
\caption{Same as Fig.~7 for intense interband exchange ($\nu_1 /\nu_0=0.6$),
amplitude and phase for $F=1$ (a,c) and $0.01$ (b,d).}
\end{figure*}

2. In spite of the existence of jumps in the behavior of individual signal
components that are associated with the vanishing of the pole due to the wave
component, the total displacement demonstrates absolutely monotonic variation
of the amplitude-phase characteristics and their derivatives. The existence
of such "compensation" within the framework of similar model was first noted
in Ref.~\onlinecite{3}. Apparently, this feature is not associated with the
particular choice of the form of the FS. We performed calculations for an FS
with different configuration -- cylinders with spherical ``caps'' at the
ends, and we obtained exactly the same result. In our opinion such ideal
``interchangeability'' attests to the fact that both components actually
describe the same process. Their separation into wave and quasiwave
components does have any profound physical meaning and is no more than a
mathematical device.

3. The phase of the transformation coefficient. From the theoretical
standpoint it should be defined as the phase of the total signal at $x_0=0$.
However, experimentally, it was determined according to the scheme arg $K =2$
arg $U(x_0) - $ arg $U(2x_0)$. For a linear phase characteristic these
definitions are equivalent, but because of the fundamentally existing
deviations from linearity the answers differ somewhat. Figure 9 shows the
phase of $K$ calculated using the experimental scheme. In the model under
discussion a sign change of the phase with increasing scattering that agrees
with the observed change (Fig.~5b) cannot be obtained for any combination of
parameters. At the same time we note that the phases of both components
forming the total signal increase with increasing scattering -- a simply
rapid drop of the amplitude of the quasiwave masks the phase increase in the
resulting signal. One possible reason for the observed discrepancy could be
that initially the magnitude of the quasiwave component is understated
because of the diffraction losses for it and for the wave component.

There arise completely natural questions of whether the FLI has an effect at
all on the process of excitation and propagation of electron sound and
whether it is possible to estimate its intensity on the basis of experiments
similar those which have been performed. The answer to the first question is
yes -- the FLI is the source of the force that supplements the forces
existing in the gas model. As a result, the stiffness of the system
increases, manifesting in the propagation velocity of the waves. Figure 10
shows the computed electron sound velocities which were obtained by
differentiating the phase of the total signal with respect to $x_0$.
Independently of the intensity of the interband exchange, the velocity of the
signals increases with increasing $F$. These changes are quite large and can
be easily recorded, but the absence of a model-independent point of reference
actually precludes the possibility of estimating $F$ from these experiments.
The possibility of separating the wave and quasi-wave components according to
the sign of the phase changes of the resulting signal when a weak transverse
magnetic field is applied was discussed in \cite{16}. From the standpoint of
the concept of the present work the result of \cite{16} signifies that for a
strong FLI the phase of the total signal should be expected to decrease and
vice versa. However, the "separation" in \cite{16} was already made at the
stage where the corresponding equations were solved, and to confirm this
result it is desirable to perform the calculations taking account of all
factors participating in the formation of the total signal.

\begin{figure}[bt]
{\epsfxsize=8.5cm\epsffile{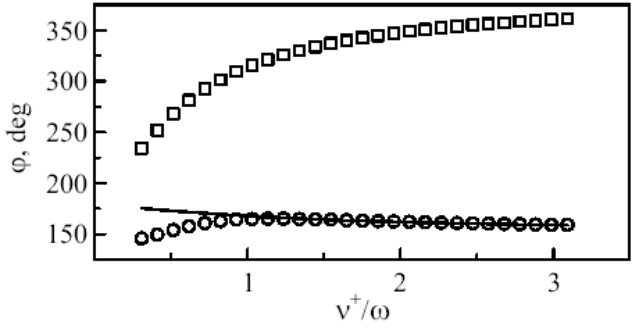}}
\caption{Computed phases of the conversion coefficients; $\nu_1 /\nu_0=0.03$,
$F =0.6$, $k_F x_0=1.2$: wave component ({\Large $\circ$}), quasiwave
component ($\square$), total signal (line).}
\end{figure}
\begin{figure}[bt]
{\epsfxsize=8.5cm\epsffile{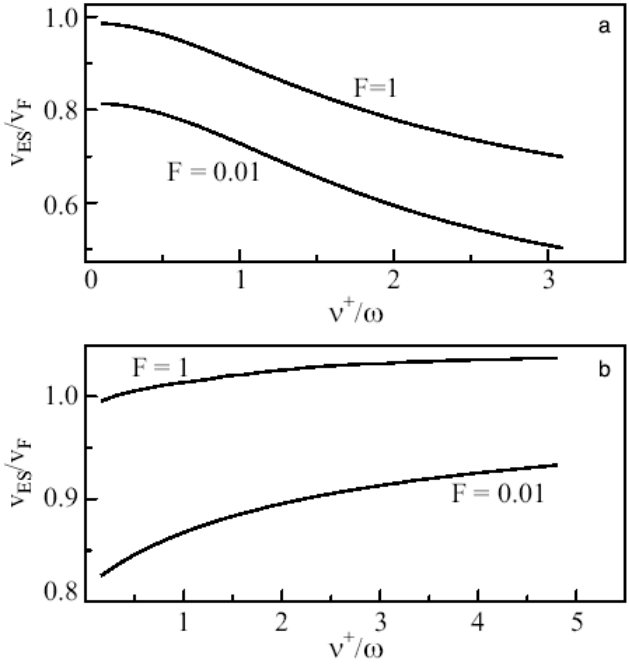}}
\caption{Effect of the FLI on the velocity of electron sound: $k_F x_0=1.2$,
$\nu_1 /\nu_0=0.03$ (a) and $\nu_1 /\nu_0=0.6$ (b).}
\end{figure}

\section{IV. BEHAVIOR OF ELECTRON SOUND AT A SUPERCONDUCTING TRANSITION}

In this very brief section we present the results of measurements of the
characteristics of the electron sound propagation in a superconductor
(Fig.~11). A completely unique conclusion follows from them -- the observed
amplitude and phase changes are independent of the thickness of the sample to
within the accuracy of the measurements, i.e., the changes concern only the
behavior of the transformation coefficient (or its square). In previous work
\cite{17} the decrease of the phase at $T_c$ was mistaken for a change of the
electron sound velocity, so that the conclusions based on this concept must
be revised. Of course, the experimentally observed linear in the energy gap
phase change will remain, but the relation of the slope of this dependence
with the FLI parameter must be determined more accurately. The evolution of
the logarithm of the modulus of the transformation coefficient appears to be
somewhat unusual for the kinetics of superconductors. The small jump is
immediately followed by a wide linear section -- as if its formation is
determined by the so-called number of superconducting electrons \cite{13}.

Likewise, the results for the geometries ${\bf q}\| [010]$ and ${\bf q}\|
[100]$ are practically identical in the superconducting phase (in contrast to
the behavior in the normal state). This means that the anomalous interband
exchange rate characteristic for the [010] axis plays no role in this case,
just like the two-band structure may not play a role either. To some extent
this makes it possible to validate the approach of Ref. \onlinecite{18},
which is based on an analysis of the single-band model neglecting the FLI and
relating the behavioral features of the electron sound below $T_c$ with the
evolution of the branch point contribution. However, even in this very simple
case it is possible to come close to a solution of the problem only in the
limit $|k_0| x_0 \gg 1$, which is far from the situation actually attainable
in experiments. For this reason, on the whole, it can be stated that the
processes determining the behavior of electron sound in a superconductor are
not clearly understood at the present time.

\begin{figure}[bt]
{\epsfxsize=8.5cm\epsffile{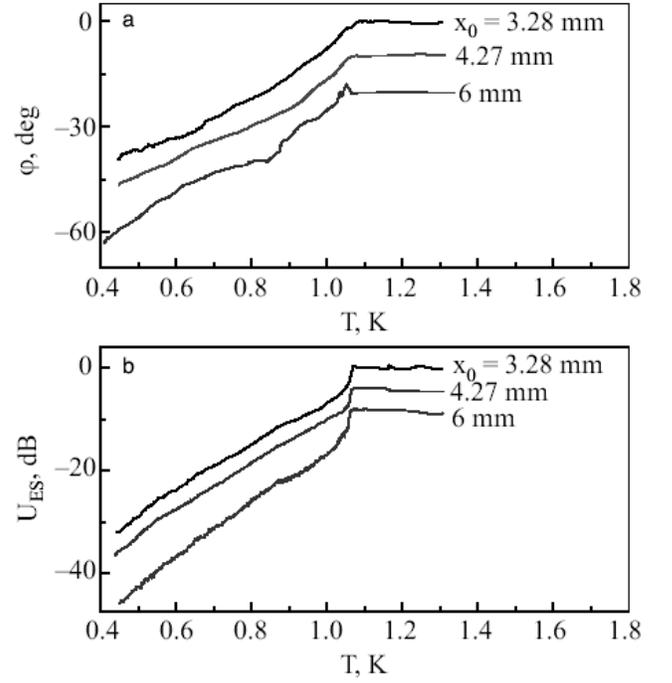}}
\caption{Effect of a superconducting transition on the amplitude-phase
characteristics of electron sound; the signal phases are shifted by 10 deg
with respect to one another (a), amplitude (4 dB shift) (b)}
\end{figure}

\section{V. CONCLUSION}

We shall now formulate the essential results obtained in this work.

1. The amplitude-phase relations characterizing the propagation of electron
sound in Ga samples of different length were studied. The modulus of the
conversion coefficient and the temperature changes of the velocity of the
wave and the phase of the transformation coefficient were determined.

2. A model problem of the excitation of electron sound in a finite-size
sample with diffuse electron scattering at the interface boundaries was
solved. It was shown that the character of the scattering has virtually no
effect on the amplitude-phase characteristics of the propagating waves.

3. It was shown that the electronic pressure at the boundary of separation
makes the main contribution to the recorded signal. As a result the amplitude
of the displacements on the receiving interface is $v_F /2s$ times greater
than that in the electron sound wave. This result is not limited to the
phenomena discussed in the present work; it is valid for any case of the
interaction of a wave, coupled with elastic deformations and moving with
supersonic velocity, with an interface.

4. It was found that the changes occurring in the amplitude and phase of an
electron sound wave in the superconducting phase are determined, first and
foremost, by the evolution of the transformation coefficient and do not
depend on the length of the sample.

We thank L.~A.~Pastur for helpful discussions and A.~I.~Petrishin for
assisting in the measurements.

\end{document}